\documentstyle[11pt,newpasp,twoside,epsf]{article} 
\begin{document} 
\title{Globular Clusters in the Central Region of Nearby Galaxy Clusters}

\author{Michael Hilker} 
\affil{Sternwarte der Universit\"at Bonn, Auf dem H\"ugel 71, 53121 Bonn, 
Germany, email: mhilker@astro.uni-bonn.de}

\begin{abstract} 

In this contribution, first results of deep VLT photometry ($V,I$) in the 
central region of the Hydra~I and Centaurus galaxy clusters are presented. In 
both galaxy clusters, many star clusters have been identified down to the
turnover magnitude of the globular cluster luminosity function at 
$V\simeq26.0$ mag. They are distributed not only around the several 
early-type galaxies, but also in the intra-cluster field, as far as 250 kpc 
from the cluster centers. Outside the bulges of the central galaxies in Hydra~I 
and Centaurus, the intra-cluster globular cluster system is dominated by blue 
clusters whose spatial distribution is similar to that of the (newly 
discovered) dwarf galaxies. 

\end{abstract}

\section{Introduction and data} 

The centers of galaxy clusters are the densest regions of galaxy populations 
in the Universe. They are the places where the most frequent interactions
between galaxies are expected to have taken place during the cluster formation
epoch (and maybe also in the present). Some striking properties of galaxy
cluster centers are: 1) a very rich globular cluster system (GCS) around the 
central galaxy (e.g. Harris 1991), 2) an extended stellar halo (cD halo) around 
the central galaxy (e.g. Schombert 1988), and 3) an abundant population
of early-type dwarf galaxies clustered towards the center (e.g. Ferguson \& 
Binggeli 1994).
How do these findings come together? Can they be the result of a
common scenario in which galaxy disruption played a major role (see Hilker et
al. 1999)? Nearby galaxy clusters provide an ideal laboratory to study the 
different stellar components in detail. 

The Hydra~I Galaxy cluster is dynamically evolved, has a regular core shape
and an isothermal X-ray gas halo that can be followed out to about 160 kpc.
The Centaurus cluster is dynamically young with two merging sub-groups, a 
main cluster component (Cen30) around the cD galaxy NGC~4696 and a smaller 
group component (Cen45) around NGC~4709.
Both galaxy clusters are located at a distance of about 45 Mpc.

Both galaxy clusters were observed at dark time and under photometric 
conditions in the filters $V$ and $I$ with FORS1 at the VLT (ESO, Paranal). 
The seeing in all fields was in the range 0.5 to 0.7 arcsec, thus providing
a very homogeneous data set.

\begin{figure}
\plotone{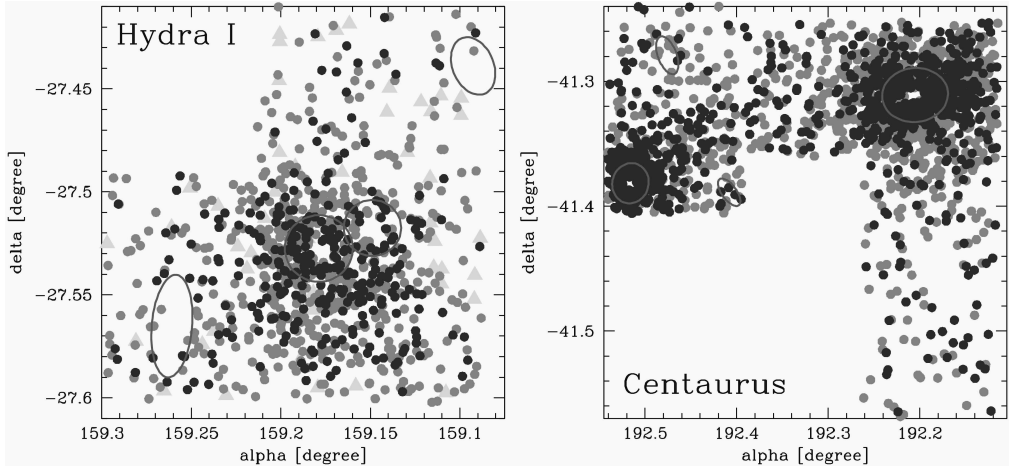}
\caption{The distribution of blue (light dots) and red (dark dots) globular 
clusters in the center of the Hydra~I (left) and Centaurus (right) cluster is 
shown. The circles indicate the location of the major galaxies, and bright 
triangles are dwarf galaxy candidates in Hydra~I.
}
\end{figure}

\section{First results and future work}

As one can see in Fig.~1, globular cluster candidates are spread all over the
core of the two galaxy clusters, well outside the tidal radii of the central
galaxies. The red GCs ($1.0<(V-I)<1.25$) are more concentrated towards the 
galaxy's
bulges than the blue GCs ($0.8<(V-I)<1.0$). The intra-cluster GCs are not 
uniformly distributed around the central galaxies. In the Centaurus cluster, 
they show a tidal tail-like structure between the two dominant giant 
ellipticals. In Hydra~I, they occupy the same space as the abundant (newly 
found) dwarf spheroidal galaxies. Also there exists a population of very blue
($0.6<(V-I)<0.8$), probably young clusters close to NGC~3311. These clusters
might have been stripped from the late-type group of galaxies around NGC~3312
that is passing by the core of Hydra~I. In both clusters, a population of
very bright ($-11<M_V<-13$) cluster or compact dwarf galaxy candidates seems
to exist.

In the near future we intend to 1) construct a density map of the 
intra-cluster GC population, in order to define its center and compare it to 
that of the X-ray gas halo, 2) model the cD halo light to study 
the local specific frequency all over the galaxy cluster, 3) study in detail 
the individual GCSs of all member galaxies down to the dwarf galaxy regime, 
and 4) confirm the membership of dwarf galaxies and bright compact objects by
follow-up spectroscopy.

\end{document}